\documentclass[aps,reprint]{revtex4-1}
\usepackage{graphicx}
\usepackage{amssymb}
\usepackage{epsfig}
\usepackage[scriptsize,bf]{caption}
\newcommand{\mev}{\mathrm{MeV}}
\newcommand{\fm}{\mathrm{fm}}

\newcommand{\br}[1]{\left(#1\right)}
\newcommand{\sq}[1]{\left[#1\right]}

\begin{document}
\title{SQM stars around pulsar PSR B1257+12}

\author{Marek Kutschera}
\altaffiliation[Also at ]{Institute of Nuclear Physics, Polish Academy of Sciences, Radzikowskego 152, PL-31342 Krak\'{o}w, Poland.}
\affiliation{Institute of Physics, Jagiellonian University, Reymonta 4, PL-30059 Krak\'{o}w, Poland.}
\author{Joanna Ja{\l}ocha}
\email{Joanna.Jalocha@ifj.edu.pl}
\affiliation{Institute of Nuclear Physics, Polish Academy of Sciences, Radzikowskego 152, PL-31342 Krak\'{o}w, Poland.}
\author{Sebastian Kubis}
\affiliation{Institute of Nuclear Physics, Polish Academy of Sciences, Radzikowskego 152, PL-31342 Krak\'{o}w, Poland.}
\author{{\L}ukasz Bratek}
\affiliation{Institute of Nuclear Physics, Polish Academy of Sciences, Radzikowskego 152, PL-31342 Krak\'{o}w, Poland.}


\begin{abstract}
\begin{description}
\item[Abstract]
Following  Wolszczan's landmark discovery of planets in orbit
 around pulsar PSR B1257+12 in 1991, over 300 planets in more than 200 planetary
 systems have been found. Therefore, the meaning of Wolszczan's discovery cannot be
 overestimated. In this paper we aim to convince the reader that the
 objects accompanying pulsar PSR B1257+12  are more exotic than thought so far.
 They might not be ordinary planets but dwarf strange quark stars,
 whereas the pulsar might be a quark star with standard mass, not a
 neutron star. If this was the case, it would indicate that strange quark matter is the ground state of
 matter.
\end{description}
\end{abstract}

\pacs{Valid PACS appear here}
                             
\keywords{Suggested keywords}

\maketitle

\section{Introduction}

The millisecond pulsar PSR B1257+12 with the estimated age of $800$ millions years, is
located at a distance of 300 parsecs from the Earth. The tiny pulsar is being
orbited by three bodies with masses $0.025$, $3.9$, and $4.3$ of the Earth mass,
typical of ordinary planets, and the respective radii of $0.19$, $0.36$ and $0.46$ $AU$
\cite{bib:wolszcz}, \cite{bib:wol}. For that reason, the bodies are considered to be planets.
It was clear from the start,  that the planetary system around PSR B1257+12 was
unusual. Before Wolszczan's discovery, the existence of planets around neutron
stars had not been considered seriously. If the planets were to be remnants of a
planetary system existing around the progenitor star of PSR B1257+12, then it would
be highly improbable for these planets to have survived the supernova explosion
accompanying the neutron star's birth,  since then compact bodies acquire
velocities high enough to escape from the binding potential of the gravitational
well of the surrounding matter. These planets might also had been formed already
after the supernova explosion and the neutron star's formation. In
particular, models of  planetary system formation around
pulsar accreting matter from its companion have been suggested  \cite{bib:banit}. However, this
scenario seems also improbable because of strong radiation in the pulsar's
vicinity \cite{bib:miller}. Therefore, another hypothesis seems worth considering.
The objects in orbit around PSR B1257+12 may be just miniature quark stars.

\section{Strange quark matter as a building material of mini-stars}

A quark star is a hypothetical object consisting of strange quark matter (SQM).
SQM is a mixture of quasi-free quarks $u$, $d$, $s$ and electrons, all of which form
relativistic fermi gas. If SQM is the ground state of strong interactions, then
one could expect strange stars to form in space. For SQM to be the ground state,
one needs the minimum of energy per baryon to be not greater than the energy per
baryon for the strongest bound nucleus in nature -- the nucleus of iron-$56$. Hence, $(\rho_{E}/n_{B})_{SQM}\leq{}E(\mathrm{Fe}^{56})/56=930.4 \mev$ at
zero pressure, where $\rho_{E}$ is the energy density and $n_{B}$ the baryon
density. The possibility of existence of such objects was put forward first
in \cite{Witten:1984rs}.
The easiest way to obtain a phenomenological equation of state for SQM, is to
consider the following simplified model  \cite{bib:chodos}. Suppose that quarks
form ultra-relativistic and electrically neutral quantum gas of non-interacting
fermions in the $\beta$-equilibrium at temperature $T=0[K]$. In this case, the
pressure $p$ and the total energy density $\rho_{E}$ are
$$p=\frac{3}{4}\br{\rho_{E}-4B},\qquad\rho_{E}=\frac{9\pi^{2/3}}{4}\hbar c
n_{B}^{4/3}+B,$$ where $n_B$ is the baryon density and $B$ is the  bag
constant, $B=50\div200\, [\mev/\fm^{3}]$. These equations are valid when the
Fermi energy is much higher than the rest energy of the heaviest quark (in our
numerical calculations we included also the strange quark mass, but the full form of
the resulting equations is not important here).

\begin{figure*}[] 
\begin{tabular}{cc} \includegraphics[width=0.4\textwidth]{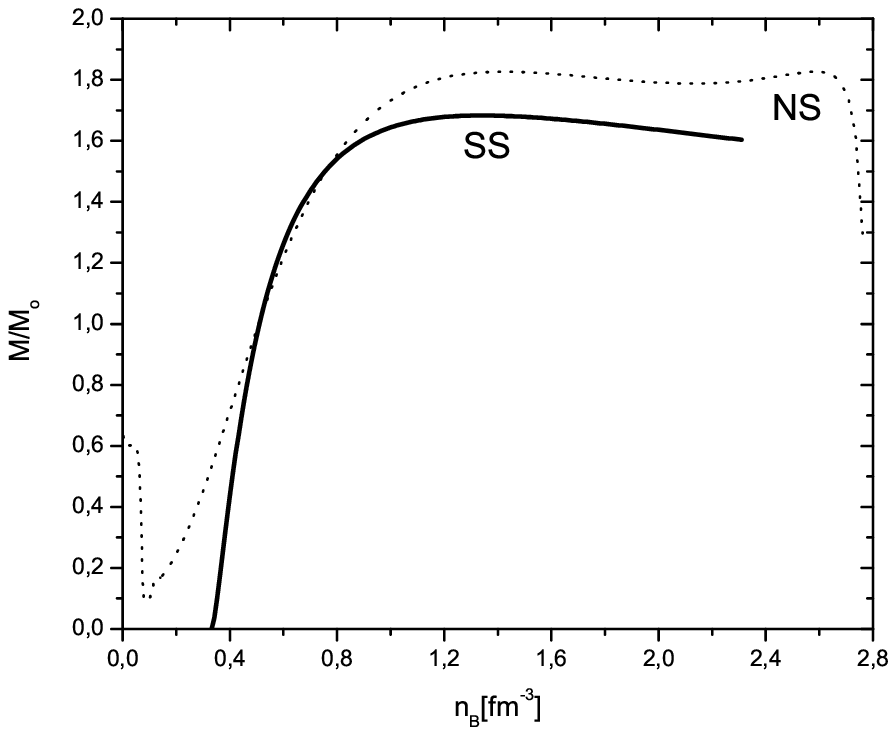}&
\includegraphics[width=0.4\textwidth]{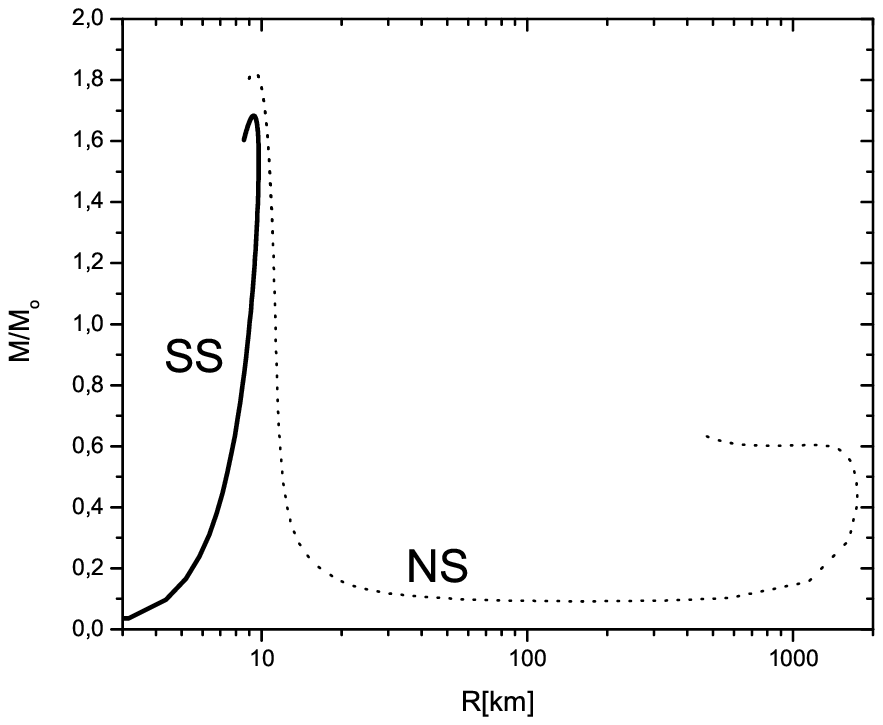}\\ (a)&(b)\\ \end{tabular}
\caption{\label{fig:1} Masses of quark stars as a function of central baryon
density compared with a similar relation for a neutron star with a crust shell
(a), and the \textit{mass-areal radius} relation for SQM stars and similar relation for a
neutron star with crust shell  (b). Both cases assume bag constant $B=70
[\mev/fm^{3}]$, the strange quark's mass $150 [\mev/c^{2}]$, and the UV14+TNI
equation of state for the neutron star.}
\end{figure*}

The maximal mass and the corresponding size of a quark star are close to those
for a neutron star.  With the assumed example values $B=70\,[\mev/\fm^{3}]$ and
the strange quark mass $m_s=150\,[\mev/c^{2}]$, SQM would be the ground
state for strong interactions. Then, the maximal mass of a quark star made of SQM
would be $\approx1.68\,M_{\odot}$ with the corresponding areal radius of
about $9.3\,[\mathrm{km}]$\footnote{One should distinguish between the areal and
physical radius of a general-relativistic star. For example, the SQM star of
maximal possible mass has its physical radius about $1.2$ times greater than its
areal radius. This distinction is usually not made in the literature and by a star's
radius is always understood its areal radius.}, 
whereas a neutron star with the
UV14+TNI equation of state \cite{Wiringa:1988tp} attains
$M_{\mathrm{max}}=1.83\,M_{\odot}$ at $R=9.5\,[\mathrm{km}]$, see figure
\ref{fig:1}.
Here, the similarities end. Only for sufficiently massive neutron stars,
their gravitational field could prevent neutron matter from disintegration.
Thus, there is the lower bound for masses of neutron stars. For realistic equations of state
the minimum is approximately $0.1 M_\odot$ \cite{Haensel:2002yz}. Unlike for neutron
stars,  there would be no such a limit for SQM stars bound mainly by quark
forces. Consequently, there are possible SQM stars of arbitrary low mass and
radius. The mass-radius diagram for SQM stars is also different, see figure
\ref{fig:1}(b). Low mass quark stars have small radii, while the radii of
neutron stars grow with decreasing mass.

Confirmation of the presence of quark stars in space would indicate that SQM is the
ground state for strong interactions.  It is thus natural that astrophysicists
have been intensively searching for such objects
\cite{bib:haensel},\cite{bib:bom}. Unfortunately, this endeavor is difficult,
since quark and neutron stars  are hardly distinguishable from each other in the range of masses
close to $1.4\,M_{\odot}$, typical of pulsars,  in which range the  mass-radius
relations are similar for both kinds of stars. On the other hand, it may turn out
that quark stars have been already  found.

\section{SQM and Wolszczan's system around PSR B1257+12}

The pulsar PSR B1257+12 is being orbited by three bodies of masses $0.025$, $3.9$
and $4.3$ in units of Earth mass. Owing to small masses, the objects are
regarded as planets. The possibility the planets might be miniature compact
stars has been already suggested in \cite{bib:xu} in the context of pulsar PSR
B1828-11 precession torqued by a quark planet. Masses of these objects are too
small for them to be neutron stars. If they were to be compact objects,
their building material would have to be made of SQM. If this really was the
case, it would mean that PSR B1257+12 must be a quark star and SQM is the ground
state for strong interactions.

\begin{figure}[]
\includegraphics[width=0.4\textwidth]{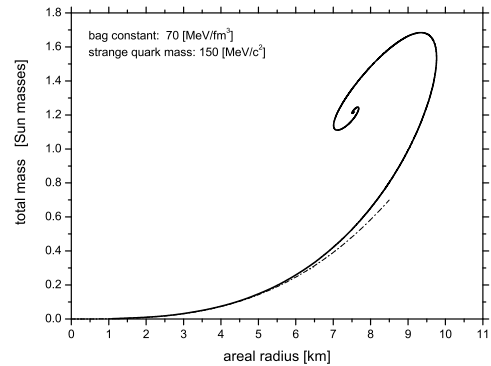}
\caption{\label{fig:mr}Comparison of Newtonian $M-R$ relation (\textit{dash-dot line}) and General-Relativistic $M-R$ relation (\textit{solid line}) for a strange quark star. The assumed strange quark mass is $150 \sq{{\mev}}$ and the bag constant is $70 \sq{\mev\cdot\fm^{-3}}$.}
\end{figure}

\begin{figure*}[]\begin{tabular}{cc}
\includegraphics[width=0.4\textwidth]{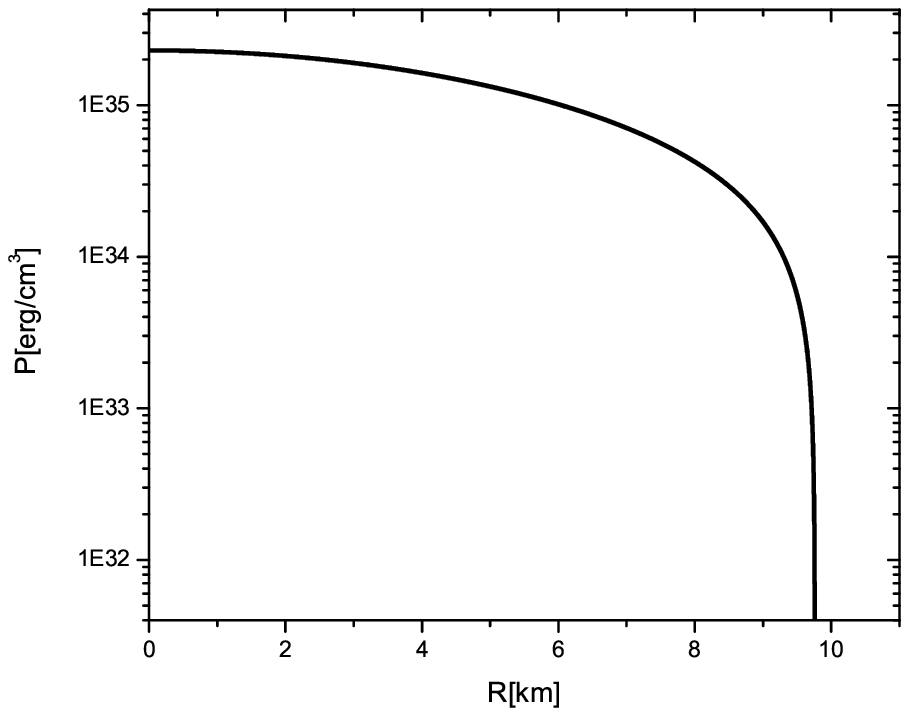}&
\includegraphics[width=0.4\textwidth]{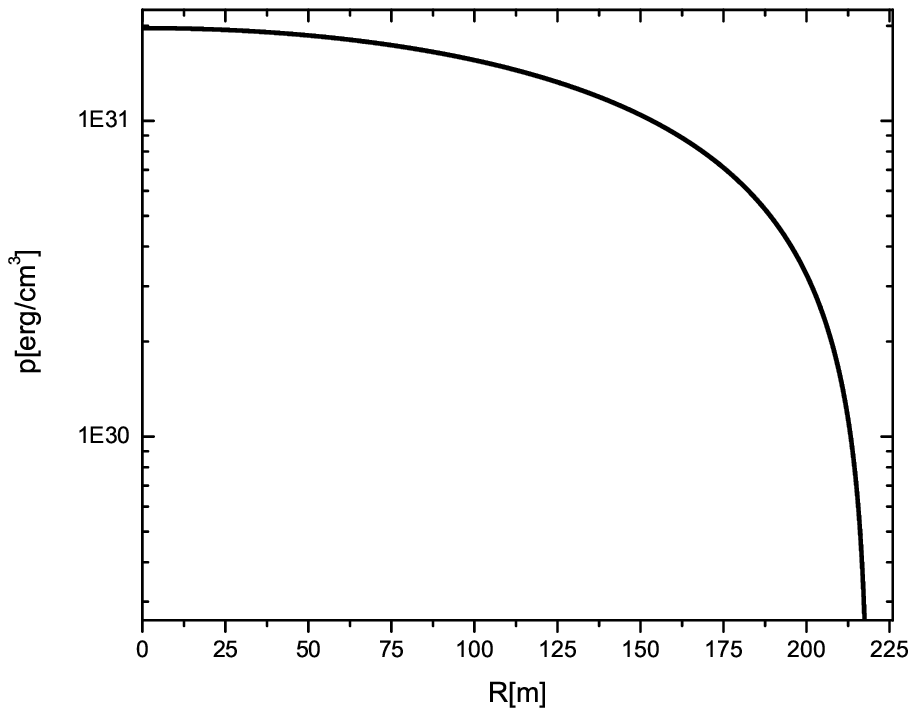}\\ a)&b) \end{tabular}
\caption{\label{fig:3} Cross section through a quark star of mass $1.4
M_{\odot}$ (a) and $3.9 M_{\oplus}$ (b). The assumed values for the bag constant
is $B=70 \mev/\fm^{3}$, and the strange quark mass $m_s=150 \mev/c^{2}$).}
\end{figure*}

Suppose that the objects orbiting PSR B1257+12 are SQM stars and assume that
$B=70\,[\mev/\fm^{3}]$ and $m_s=150\,[\mev/c^{2}]$. Since for such small stars
the energy density must tend to the energy density on their surface
$\rho_{ext}$, and since $2GM(r)/(r c^2)\ll1$, the relation between the total
mass $M$ and the physical radius $R$ of a low mass SQM star is nothing but
$M=\frac{4\pi}{3}\rho_{ext}R^3$ which, for the assumed parameters, gives the following
mass-radius relation in Newtonian limit $M
R^{-3}\approx1.1413\times10^{-3}
[\mathrm{M}_{\odot}/\mathrm{km}^3]$\footnote{At a given $B$, $\rho_{ext}$ is
independent of the total mass. We calculated $\rho_{ext}$ at vanishing pressure
using a relativistic equation of state in which low mass quarks and electrons
are considered massless.} (comparison between this relation and that
obtained in the framework of General Relativity is illustrated in figure \ref{fig:mr}).
Therefore, the stars' radii are of about
$40$, $217$, and $225$ meters, respectively. The central pulsar of mass $1.4
M_{\odot}$ has the areal radius of about $9.75\,[\mathrm{km}]$. Figure \ref{fig:3}
shows energy density and pressure as functions of the areal radius for two example
SQM stars.

What could have caused the mini-quark stars to find themselves in the pulsar vicinity? The first possibility is detachment of quark matter from the central pulsar during its formation phase, when the "boiling" strange matter core could eject some fragments, so that they start to orbit around the strange star. However, this mechanism does not explain why the orbits of the three planets are nearly circular and almost coplanar (at least, for
two of them) \cite{Konacki:2003xa}.  These features make the planetary disk origin of the pulsar planets very likely. The planetary disk stability and
the planet formation present a lot of unknowns and require ad hoc assumptions \cite{phinney:1993}.

Here we propose another scenario.
The present rotation period of pulsar B1257+12 is of about $P=6.2$ ms and it
decelerates with the rate $\dot{P}= 1.14\times 10^{-19}$
\cite{Manchester:2004bp}. The pulsar has the feature of a pulsar spun up in
a binary system -- it is a millisecond pulsar and, at the same time, it has
comparably small deceleration rate and weak magnetic field. It seems therefore
probable the pulsar might have had a companion from which it had been accreting
matter. In this case it is difficult to estimate the age and the initial rotation velocity
of a compact object. However, during the accretion phase, the compact object, by
assumption being a quark star, could have spun up to the velocities necessary to detach
matter from its surface in the equatorial region. If only the accretion were sufficient to
accelerate the star to the Keplerian frequency, which is about 1 KHz for strange star
\cite{Haensel:2009wa}, then several bubbles of quark matter could
start to orbit the central star. In a natural way their trajectories would be
circular and coplanar. Being pieces of strange matter they would be
completely  resistant to the unfriendly pulsar environment
heated by pulsar wind or accretion which may evaporate the
normal planet.  

\section{Summary}
In this paper we put forward the hypothesis that the planets
orbiting the pulsar PSR B1257+12 may be quark stars made of strange quark matter.
This is alternative to the currently accepted hypothesis that the system is
composed of "ordinary" planets.

\nopagebreak

\end{document}